\documentstyle[12pt]{article}
\begin{document}
\baselineskip .3in
\pagestyle{plain}
\newpage
\begin{center}
{\Large{\bf Two-site polaron problem : a perturbation approach 
with variational basis states}}\\
\end{center}
\vskip 1.0cm
\begin{center}
 A. N. Das\footnote{ e-mail: atin@cmp.saha.ernet.in} and Jayita 
 Chatterjee\footnote{ e-mail: moon@cmp.saha.ernet.in} 
\end{center}
\vskip 0.50cm
\begin{center}

 {\em Saha Institute of Nuclear Physics \\
1/AF Bidhannagar, Calcutta 700064, India}\\

\end{center}

\vskip 1.0cm

PACS No.71.38. +i, 63.20.kr  
\vskip 1.0cm
\newpage
\begin{center}
{\bf Abstract}
\end{center}
\vskip 0.5cm

A convergent perturbation method using modified Lang Firsov 
transformation is developed for a two-site single-polaron system. 
The method is applicable for the entire range of the electron-phonon 
coupling strength from the antiadiabatic limit to the intermediate 
region of hopping. The single-electron energies, oscillator wave 
functions and correlation functions, calculated using this method, 
are in good agreement with the exact results.

\newpage
%\section{Introduction}
%\label{sec:1.}

%\sloppy

%\pagestyle{plain}
\begin{center}
{\bf 1. Introduction}   
\end{center}
\vskip 0.3cm 

The interaction of conduction electrons with lattice vibrations 
is described by the so called electron-phonon problem. 
The Holstein model \cite {Hol} is one of the fundamental models 
which has been studied widely in this context. The model consists 
of a one-electron hopping term, Einstein phonons at each site 
and a site-diagonal interaction term which couples the electron 
density and ionic displacements at a given site. For weak 
electron-phonon (e-ph) coupling the frequency of the phonons and the 
effective mass of the the electron are renormalized, which are 
described by the Migdal approximation \cite{Mig}. For large e-ph 
coupling the electrons are self-trapped in the lattice 
deformation producing a small polaron. The motion of the electron is then 
accompanied by the lattice deformation. This results in a large 
effective mass or reduced effective hopping of the dressed electrons 
(polarons). 
The Lang-Firsov (LF) method based on the LF canonical \cite {LF} 
transformation works in this strong coupling region in the 
antiadiabatic limit. For weak coupling the Migdal approximation is 
satisfactory. However, no conventional analytical method exists 
at present which is beleived to describe a Holstein model for the 
entire range of the coupling strength. So, it would be useful to 
develop or identify an analytic method which could be applied to 
both the strong and weak coupling cases.
Ranninger and Thibblin \cite {RT} 
made an exact diagonalization study of a two-site polaron problem 
and showed that the behavior of the polaron differs very much from 
that predicted by the classical LF method. Marsiglio \cite {Mar} 
extended those calculations to the bulk limit in one dimension 
by studying the Holstein model with one electron up to 16 site 
lattices. He concluded that for intermediate coupling strength
neither the Migdal nor the small-polaron approximation is in 
quantitative agreement with the exact results. Kabanov and 
Ray \cite{KR} and Alexandrov $et~al.$ \cite {Alex} noted that   
for $t> \omega_0$ the adiabatic small-polaron 
approximation describes the ground state 
energy accurately except for intermediate 
coupling strength. Ranninger and Thibblin 
\cite{RT} and Marsiglio \cite{Mar} studied also the correlation 
functions using exact diagonalization technique with the finite size 
Holstein model and found that the results are non trivial and cannot 
be described by any conventional analytical method. Recently, 
de Mello and Ranninger \cite{MR} emphasized this point further.

The objective of the present work is to search for an analytical 
method which may be applicable reasonably well for the major range 
of e-ph coupling strength. 
For our study we consider a two-site one-electron Holstein model 
for which 
exact results are available \cite {RT}. Previously, we \cite {DC, CD} 
investigated the ground state energy and the nature of polarons 
in a two-site and a four-site Holstein model using the modified 
Lang Firsov (modified LF) transformation and two-phonon coherent 
states and 
found that the energy obtained within such method is very close to 
the exact result. In this work we develop a perturbation expansion 
within the modified LF transformation and show that this expansion 
converges for the entire range of the coupling strength from the 
antiadiabatic limit to the intermediate region of hopping. 
The energy and the correlation functions 
calculated within our approach are almost identical with the 
exact results.

\vskip 1.0cm

\begin{center}
{\bf 2. Formalism }
\end{center}
\vskip 0.3cm

The Hamiltonian of a two-site one-electron Holstein model reads as
\begin{eqnarray}
H = \sum_{i,\sigma} \epsilon n_{i \sigma} - \sum_{\sigma}
t (c_{1 \sigma}^{\dag} c_{2 \sigma} + c_{2 \sigma}^{\dag} c_{1 \sigma})
+ g \omega_0  \sum_{i,\sigma}  n_{i \sigma} (b_i + b_i^{\dag}) 
+  \omega_0 \sum_{i}  b_i^{\dag} b_i 
\end{eqnarray} 
where $i$ =1 or 2, denotes the site. $c_{i\sigma}$ ($c_{i\sigma}^{\dag}$)  
is the annihilation (creation) operator for the electron with spin 
$\sigma$ at site $i$ and $n_{i \sigma}$ (=$c_{i\sigma}^{\dag} c_{i\sigma}$) 
is the corresponding number operator, $g$ denotes the e-ph coupling  
strength, $b_i$ and $b_{i}^{\dag}$ are the annihilation and 
creation operators, respectively, for the phonons corresponding to 
interatomic vibrations at site $i$, $\omega_0$ is the phonon frequency. 
In Hamiltonian (1) there is no spin dependent or spin reversal term 
so for the study of one-electron case the spin index is 
redundant. In the following we shall not use the spin index.

Introducing new phonon operators $a=~(b_1+b_2)/ \sqrt 2$ and 
$d=~(b_1-b_2)/\sqrt 2 $ the Hamiltonian is separated into two parts   
($H=H_d + H_a)$ : 
\begin{eqnarray}
H_d = \sum_{i} \epsilon n_{i} - t (c_{1}^{\dag} c_{2} + c_{2}^{\dag} c_{1}) 
+ \omega_0  g_{+} (n_1-n_2) (d + d^{\dag}) 
+  \omega_0  d^{\dag} d 
\end{eqnarray}
and  
\begin{equation}
H_a =  \omega_0 \tilde{a}^{\dag}\tilde{a} - \omega_0 n^2 g_{+}^2 
\end{equation}
where $g_{+}=g/\sqrt 2$, $\tilde{a}=a +ng_{+}$ and 
$\tilde{a}^{\dag}=a^{\dag} +ng_{+}$.

$H_a$ describes a shifted oscillator which couples only  
with the total number of electrons $n(=n_1+n_2)$, which 
is a constant of motion. The last term 
in Eq.(3) represents lowering of energy achieved through 
the lattice deformations of sites 1 and 2 by the total number 
of electrons. 

$H_d$ represents an effective e-ph system where phonons 
directly couple with the electronic degrees of freedom. 
$H_d$ cannot be solved exactly by any analytical method.  
We now use the modified LF transformation where the lattice 
deformations are treated as variational parameters 
\cite {DC, DS, LS}. For the present system, 
\begin{equation}
\tilde{H_d} = e^R H_d e^{-R}
\end{equation}
where $R =\lambda (n_1-n_2) ( d^{\dag}-d)$, $\lambda$ is a
variational parameter and linearly related to the displacement of 
the d oscillator.

The transformed Hamiltonian is then obtained as 
\begin{eqnarray}
\tilde{H_d} &=&  \omega_0  d^{\dag} d + \sum_{i} \epsilon_p n_{i} - 
t [c_{1}^{\dag} c_{2}~ \rm{exp}(2 \lambda (d^{\dag}-d))    \nonumber\\ 
&+& c_{2}^{\dag} c_{1}~\rm{exp}(-2 \lambda (d^{\dag}-d))]  
+ \omega_0  (g_{+} -\lambda) (n_1-n_2) (d + d^{\dag}) 
\end{eqnarray}
where 
\begin{equation}
\epsilon_p = \epsilon - \omega_0 ( 2 g_{+} - \lambda) \lambda 
\end{equation}

It may be mentioned that with an ordinary LF transformation, 
where one chooses a phonon basis of the oscillator with a fixed 
displacement ($\lambda= g_{+}$ for this case), one can diagonalize 
the Hamiltonian in absence of hopping. The hopping term, containing 
off-diagonal matrix elements in the new phonon basis, may then be 
treated within the perturbation approach in the strong coupling 
and antiadiabatic limit \cite{FK}. However, in order to develop a 
perturbation theory to be valid for the entire range of coupling strength 
one should consider a variational phonon basis such that the major part of 
the hamiltonian could be diagonalized for different values of the coupling 
strength. The modified LF transformation, where the phonon basis are formed by 
the oscillator with variable displacement, would serve this purpose. 

For the single polaron problem we choose the basis set (for $\tilde{H_d}$)
\begin{eqnarray}
|+,N \rangle = \frac{1}{\sqrt 2} (c_{1}^{\dag} + c_{2}^{\dag}) 
|0\rangle_e  |N\rangle \nonumber\\   \\
|-,N \rangle = \frac{1}{\sqrt 2} (c_{1}^{\dag} - c_{2}^{\dag}) 
|0\rangle_e  |N\rangle \nonumber
\end{eqnarray}
where $|+\rangle$ and $|-\rangle$ are the bonding and antibonding 
electronic states and $|N\rangle$ denotes the $N$th excited oscillator  
state. 

Note that the last term in Eq.(5) has only off-diagonal matrix 
elements connecting bonding and antibonding states with the change in
phonon number by $\pm 1$,
\begin{eqnarray}
\langle M , \pm| \omega_0  (g_{+} -\lambda) (n_1-n_2) (d + d^{\dag}) 
|\mp , N \rangle \nonumber \\
= (\sqrt N~ \delta_{M,N-1} +\sqrt{N+1}~ \delta_{M,N+1})\omega_0(g_{+}-\lambda)
\end{eqnarray}
while the hopping term $H_t$=  
$- t [c_{1}^{\dag} c_{2}~ \rm{exp}(2 \lambda (d^{\dag}-d)) + c_{2}^{\dag} 
c_{1}~\rm{exp}(-2 \lambda (d^{\dag}-d))]$
has both diagonal and off-diagonal elements in the chosen basis.
The diagonal part of $H_t$ is given by,
\begin{eqnarray}  
\langle N, \pm|H_t|\pm , N\rangle=\mp t_e\sum_{i=0}^{N}\left[
\frac{(2\lambda)^{2i}}{i!}(-1)^i N_{C_i}\right]
\end{eqnarray}  
where $t_e=t~\rm{ exp}{(-2\lambda^2)}$ and $N_{C_i}=\frac{N!}{i!(N-i)!}$.
              
The diagonal part of the Hamiltonian $\tilde{H_d}$ (in the chosen 
basis) is considered as the unperturbed Hamiltonian ($H_0$) and the 
remaining part of the Hamiltonian $H_{1}= \tilde{H_{d}}-H_0$ is 
treated as a perturbation. 

The unperturbed energy of the state $| \pm,N\rangle$is given by 
\begin{eqnarray}
 E_{\pm,N}^{(0)}= \langle N,\pm|H_0|\pm, N \rangle=
 N\omega_0 +\epsilon_p \mp t_e \left[ \sum_{i=0}^{N}
 (\frac{(2\lambda)^{2i}}{i!} (-1)^i N_{C_i}\right]
 \end{eqnarray}
The general off-diagonal matrix elements of 
$H_1$ between the two states $|\pm,N \rangle$ and $|\pm,M \rangle$ 
are calculated as (for $(N-M)>0$)
\begin{eqnarray}  
\langle N,\pm|H_1|\pm, M \rangle &=& P(N,M)~~~\rm{for}~\rm{even}~(N-M) \\ 
 \nonumber \\
\langle N,\pm|H_1|\mp,M \rangle &=& P(N,M)+ \sqrt{N}\omega_0 
(g_{+}-\lambda)  \delta_{N,M+1}~~
\rm{for}~\rm{odd}~ (N-M). 
\end{eqnarray}\\  
where 
\begin{eqnarray}  
P(N,M) = \mp t_e (2\lambda)^{N-M} 
\sqrt{\frac{N!}{M!}} \left[  \frac{1}{(N-M)!}+\sum_{R=1}^{M}
[(-1)^R \right. \nonumber \\
 \left. \frac{(2\lambda)^
{2R}}{(N-M+R)! R!}M(M-1)...(M-R+1) ] \right] \nonumber 
\end{eqnarray}  
In the following we present the perturbation corrections 
to the energies and the correlation functions for the ground 
and the first excited state.

\vskip 1.0cm
\begin{center}
{\bf 3. The energies and the correlation functions}
\end{center}
\vskip 0.3cm
\it{A.} Ground state :

\rm  For the system considered, the state $|+\rangle|0\rangle$ has
the lowest unperturbed energy, $ E_0^{(0)}=\epsilon_p-t_e$. 
The matrix element connecting this 
ground state and an excited state $|e,N\rangle$ is given by
\begin{eqnarray}  
\langle N,e|H_{1}|+,0\rangle&=&\left[ -t_e\frac{(2\lambda)^N}
{\sqrt{N!}}+\omega_0(g_{+}-\lambda)\delta_{N,1} \right] \delta_{e,-} 
\hspace{.4cm} \rm{for~odd ~ N }\nonumber \\
&=& \left[ -t_e\frac{(2\lambda)^N}{\sqrt{N!}} \right] \delta_{e,+} 
\hspace{.4cm}    \rm{for~ even~ N }
\end{eqnarray}  
 
The first order correction to the ground state wave function is 
obtained as,

\begin{eqnarray}  
|\psi_0^{(1)}\rangle &=&-\frac{\omega_0(g_{+}-\lambda)-2\lambda t_e}
{[\omega_0+2t_e(1-2\lambda^2)]}~|-,1\rangle +\sum_{N=2,4,..}
\frac{t_e(2\lambda)^N}{\sqrt{N!} (E_{+,N}^{(0)}- E_0^{(0)})}~|+,N\rangle \nonumber\\
&+& \sum_{N=3,5,..}\frac{t_e(2\lambda)^N}{\sqrt{N!}(E_{-,N}^{(0)}-
E_0^{(0)})}~|-,N\rangle 
\end{eqnarray}  
where $E_{\pm,N}^{(0)}$ is the unperturbed energy of the state 
$|\pm,N\rangle$ as given in Eq.(10).

The second order correction to the ground state energy is given by
\begin{eqnarray}
E_0^{(2)} &=& -\frac{t_e^2(2\lambda-\frac{\omega_0}{t_e}(g_{+}-
\lambda))^2}  {\left[\omega_0+2t_e(1-2\lambda^2)\right]} 
-\sum_{N=2}^{\infty}\frac{t_e^2 (2\lambda)^{2N}}
{N!(E_{e,N}^{(0)}-E_0^{(0)})} 
\end{eqnarray}
where e=+ or - for even and odd N, respectively.

Now, one has to make a proper choice of $\lambda$, hence choice for 
the displaced phonon basis, so that the perturbative expansion becomes 
convergent. In the usual modified LF method $\lambda$ is found 
out by minimizing the ground state energy of the system. 
Here we adopt that method and check whether it gives satisfactory 
results. From our previous studies \cite {DC,CD,DS} we know that 
$\lambda$ remains small as long as $g_{+}<1$, while for large values 
of $g_{+}$ (in the strong coupling limit) it approaches or attains 
the full LF value of $g_{+}$ (see Table I). 
For small values of $\lambda$ 
the perturbation series involving linear Frohlich type (polaron-phonon) 
interaction term ($\propto (g_{+}-\lambda)$) converges automatically, 
while that involving hopping, containing powers of $2 \lambda$, 
would converge provided $t<\omega_0$ or $t\sim$ $\omega_0$. 
For strong coupling, as $\lambda$ approaches to $g_{+}$, the Frohlich 
(polaron-phonon) interaction term almost vanishes as well as $t_e$ 
becomes very small so the 
perturbation series converges. Thus, it is expected that the 
perturbation method following the modified LF transformation would 
work satisfactorily in both the weak and strong coupling limits. 
In this work we have shown that it works reasonably well for  
whole range of the coupling strength for $t \le \omega_0$. 

Following the spirit of the modified LF method the value of $\lambda$ 
is found out as $\lambda=\omega_0g_{+}/(\omega_0+2t_e)$
from minimization of the unperturbed ground state energy. It is 
interesting to note that for this particular choice of $\lambda$, 
the coefficient of $|-,1\rangle$ in Eq. (14) as well as 
the first term in the r. h. s. of Eq. (15) vanishes. 
In other words, the off-diagonal matrix element between the 
states $|+,0\rangle$ and $|-,1\rangle$ becomes zero for the modified LF 
choice of phonon basis which leads to small perturbation 
correction to the ground state within this method.

To check whether the perturbation series is converging properly 
we have calculated and computed the third order correction to 
the energy for the ground state. 
The third order correction ($ E_0^{(3)}$) to the ground state 
energy is given by,

\begin{eqnarray}
E_0^{(3)} &=& \sum_{k\neq 0}[ (H_1)_{0k}\sum_{m\neq 0}[ \frac{(H_1)_{km}
(H_1)_{m0}}{(E_{0}^{(0)}-E_k^{(0)})(E_{0}^{(0)}-E_m^{(0)})} ]~]
\end{eqnarray}
where the subscript k,m denote the states $|\pm,N\rangle$ with the 
unperturbed energy 
$E_{\pm,N}^{(0)}$ and the subscript 0 refers to the ground state 
$|+,0\rangle$. The off-diagonal matrix elements of $H_1$
are calculated using Eqs.(11) and (12).
\par The ground state wave function of $\tilde{H_d}$ considering up to 
the second order corrections in perturbation is given by,
\begin{eqnarray}  
|\tilde {G} \rangle  \equiv |+,0\rangle+|\psi_0^{(1)}
\rangle + |\psi_0^{(2)}\rangle \nonumber
\end{eqnarray} \\
which can be alternatively written as,
\begin{eqnarray}  
|\tilde{G} \rangle = |+,0\rangle +\sum_{N=2,4,..}
a_{N}|+,N \rangle +\sum_{N=1,3,..}b_{N} |-,N \rangle 
\end{eqnarray} \\
The coefficients $a_{N}$ and $b_{N}$ are determined from Eq.(14) and 
the second order correction to the wave function.
The normalized ground state wave function $|\tilde{G} \rangle_{N}$ is 
$$|\tilde{G} \rangle_{N}=\frac{1}{\sqrt{N_G}}|\tilde{G} \rangle$$ 
where $N_G$ is obtained as
\begin{eqnarray}  
N_G \equiv \langle \tilde{G}|\tilde{G}\rangle =1
+\sum_{N=2,4,..}a_{N}^2+\sum_{N=1,3..}b_{N}^2 
\end{eqnarray} \\
Within the modified LF method the ground state wave function for the 
$d$ oscillators is a displaced Gaussian
\begin{eqnarray}  
\phi(x)=\frac{1}{\pi^{\frac{1}{4}}}\rm{exp}[-(x-x_0)^2] 
\end{eqnarray}  
where, $x_{0}= - (n_1-n_2) \sqrt{2}~\lambda$.
Including the corrections due to the perturbation 
the ground state wave function for the $d$ oscillator  
is obtained as,
\begin{eqnarray}  
G(x)&\equiv& \tilde{G}(x-x_{0})= \langle x-x_{0}|0\rangle +\sum_{N=2,4..}a_N 
\langle x-x_{0}|N\rangle \nonumber\\
&+& \sum_{N=1,3...}b_N\langle x-x_{0}|N\rangle 
\end{eqnarray}  
Note that $G(x)$ and $\tilde{G}(x)$ are the 
ground state oscillator wave functions for $H_d$ and $\tilde{H_d}$, 
respectively. If the electron is located on site 1 then 
$x_{0}= -\sqrt{2}~\lambda$. 

\begin{itemize}
\item Correlation function calculation:
\end{itemize}

The static correlation function $\langle n_1 u_{1}\rangle_{0}$ and 
$\langle n_1 u_{2}\rangle_{0}$,
where $u_1$ and $u_2$ are the lattice deformations at site 1 and 2  
respectively, produced by an electron at site 1, 
indicates the strength of polaron induced lattice deformation and their 
spread. These correlation functions are determined as 
\begin{eqnarray}  
\langle n_{1} u_{1} \rangle_{0}&=&\frac{1}{2}
\left[-(g_{+} +\lambda) + \frac{A_0}{N_G}\right] \\
\langle n_{1} u_{2} \rangle_{0}&=&\frac{1}{2}
\left[-(g_{+}-\lambda)- \frac{A_0}{N_G}\right] \nonumber
\end{eqnarray}\\  
where 
\begin{eqnarray} 
A_0\equiv\langle \tilde{G} |n_1(d+d^{\dag})|\tilde{G}\rangle=\sum_
{N=1,3,..}b_N\left[
\sqrt{N}~a_{N-1}+\sqrt{N+1}~a_{N+1}\right] \nonumber
\end{eqnarray} 

\vskip 0.7 cm
{\it B. The First Excited State:}
\vskip 0.3 cm

The unperturbed energies of the states $|+,1\rangle$ and 
$|-,0\rangle$ are ($\epsilon_p +\omega_0 -t_e(1-4\lambda^2)$) and 
($\epsilon_p +t_e$), respectively. For $2t > \omega_0$, the energy 
of the state $|+,1\rangle$ is lower than that of $|-,0\rangle$ 
for $g_{+}=0$, while it is higher for large values of $g_{+}$ when $t_e$ 
becomes negligible. The off-diagonal matrix element of $\tilde{H_{d}}$ 
between these two states is nonzero. Crossing of the unperturbed 
energies of these two states at an intermediate value of $g_{+}$ and 
 nonzero off-diagonal matrix elements requires that one should 
 follow the degenerate perturbation theory. So, linear combinations 
 of the states $|+,1\rangle$ and $|-,0\rangle$ are formed to obtain 
 two new elements of basis states so that $\tilde{H_{d}}$ becomes 
 diagonal in the sub-space spanned by these two states.
 The first excited state of $\tilde{H_{d}}$ is described by one of the linear 
 combinations which has lower energy. The unperturbed first excited 
 state is given by 
\begin{eqnarray}  
|\psi_1^{(0)}\rangle= a |-,0\rangle  + b|+,1\rangle
\end{eqnarray}  
The ratio ($c$) of the coefficients $a$ and $b$ and the unperturbed energy  
($\alpha$) of the first excited state may be found out from the relation 
\begin{eqnarray}  
c = \frac{\alpha-H_{11}}{H_{12}}
=\frac{H_{12}}{\alpha-H_{22}}
\end{eqnarray}  
where $H_{11}$, $H_{22}$, $H_{12}$ are the matrix elements of $\tilde{H_{d}}$
 in the subspace of $|-,0\rangle$ and 
$|+,1\rangle$ and are given in the matrix form in the following,

\begin{eqnarray} 
\begin{tabular}{c|c c}
{$\tilde{H_{d}}$} & {$|-,0\rangle$} & {$|+,1\rangle$}\\ \hline\\
{$\langle 0,-|$} & {$(\epsilon_p+t_e)$} & 
{$2\lambda t_e+\omega_0(g_{+}-\lambda)$}\\ \\
{$\langle 1,+|$} & {$2\lambda t_e+\omega_0(g_{+}-\lambda)$} & 
{$\omega_0+\epsilon_p-t_e(1-4\lambda^2)$}\\
\end{tabular}  
\end{eqnarray}  

Eq.(23) gives two roots of $\alpha$, the lower value of $\alpha$ 
(say, $\alpha_1$) corresponds to the first excited state. 

The first order correction to the first excited state wave function 
is obtained as,
\begin{eqnarray}  
|\psi_1^{(1)}\rangle &=&\frac{1}{\sqrt{1+c^2}} \left[ \sum_{N=2,4,..}
\frac{W_e}{(\alpha_1-E_{-,N}^{(0)})}~~ |-,N\rangle \right. \nonumber\\
&+&\left. \sum_{N=3,5..}\frac{W_o}{( \alpha_1-E_{+,N}^{(0)})}
~~|+,N\rangle  \right]
\end{eqnarray}  
where $W_o=t_e\frac{(2\lambda)^N}{\sqrt{N!}}(1+2\lambda c-\frac{cN}
{2\lambda})$ \\
and $W_e= W_o +\sqrt{2} \omega_0 c(g_{+}-\lambda)\delta_{N,2}$

Second order correction to the first excited state energy is given by,
\begin{eqnarray}
E_0^{(2)} = \frac{1}{1+c^2}\left[ \sum_{N=2,4,..}
\frac{|W_e|^2}{(\alpha_1-E_{-,N}^{(0)})} 
+ \sum_{N=3,5..}\frac{|W_o|^2}{( \alpha_1-E_{+,N}^{(0)})}  \right]
\end{eqnarray}

\vskip 1.0cm
\begin{center}
{\bf 4. Results and discussions}
\end{center}
\vskip 0.3cm

In this paper we report mainly the results of $t$=1.1 (in a scale of 
$\omega_0$ =1) for which exact results \cite{RT} are available. 
For the ground state the wave function and the energy have been 
calculated up to the second order and the third order in perturbation,  
respectively. For the numerical calculation we consider up to 25 
phonon states in the series of Eqs. (14), (15) and (16). It is found 
that except 
for very high values of $g_{+}$ cosideration of 20 phonon states 
is more than sufficient, while for large values of $g_{+}$ (1.8-2.2) 
consideration of 25 phonon states is enough. 

In Table-I we have shown the unperturbed energy, the second and third 
order corrections to the ground state energy. It is seen that 
the magnitude of the higher order corrections decreases rapidly 
which clearly indicates the convergence of the series and 
reasonability of our approach. The second and third order perturbation 
corrections to the energy are small in both the weak and strong coupling 
limits and appreciable only in the intermediate coupling limit 
$1.0\leq g_+\leq 1.3$ where higher order corrections may be necessary.

It may be noted that for lower values of $t$ (results for $t=1.1$ 
are shown here) the perturbation series converges more rapidly with 
smaller perturbation corrections. So, the present method based on 
modified LF transformation is expected to work very satisfactorily 
for $t < \omega_0$. 

In Fig. 1 we have shown the single electron energies as a function 
of $g_{+}$ for the ground state (calculated up to the third order) 
and the first excited state (calculated up to the second order). 
The results are found to be almost identical with the exact results 
by Ranninger and Thibblin (within the resolution of Fig. 1 of Ref. 4). 
It should 
be mentioned that in a range $1.2<g_{+}<1.4$ the second order 
correction to the energy for the first excited state is 
$\sim 10-12\%$ of the unperturbed energy and so third order  
correction may be necessary in this region. For other 
regions the second order correction to the energy of the first 
excited state is small. 

In Fig. 2 we have shown the ground state wavefunction for the $d$ 
oscillator as a function of position $x$ for different values of 
the e-ph coupling when the electron is located on site 1. 
For weak coupling ($g_{+}<1)$ the wave function shows displaced  
Gaussian like single peak where the displacement is given by the 
modified LF value, $x_{0}= - \sqrt{2}~\lambda$. However, for $g_{+}$=1.3 
an additional prominent shoulder appears. For higher values of $g_{+}$ 
this shoulder 
takes the form of a broad peak. These results are completely consistent 
with the results obtained by Ranninger and Thibblin by exact 
diagonalization study \cite{RT}. 

In Fig. 3 we have plotted the variation of the correlation functions 
$\langle n_{1} u_{1} \rangle_{0}$ and $\langle n_{1} u_{2} \rangle_{0}$  
with $g_{+}$. Our perturbation results are found to  
be very close to the exact results of 
Ref. \cite{RT}. It may be mentioned that the second order correction 
to the ground state wave function becomes very important in 
determining the shape of the correlation function within our method. 
In our method the ground state has no component of 
$| -,1\rangle$ up to the first order correction, but it appears 
in the second order correction to the wavefunction. Presence of 
$|-,1\rangle$ in the ground state has a significant contribution 
to the correlation function. 
The correlation functions are found to be very sensitive to any small 
correction to the wave function unlike the single elctron energies. 
We find a slight departure of our results from the exact results
of Ref. cite{RT} at intermediate coupling strength. This is due to
the finite series (up to the second order in perturbation for the
wave function) that we have considered. 

It may be mentioned that recently de Mello and Ranninger concluded from 
their study of a two site one polaron problem that it would be very difficult 
for any analytical method to describe the Holstein model except in the extreme 
adiabatic or nonadiabatic limit. We have shown here that even for 
$t/\omega_0=1.1$ which is in between the above two limits the exact results 
are fairly reproducible by our analytical method based on the modified LF 
transformation and perturbation expansion. Comparisons of the energies and 
the correlation functions in the ground state with exact results show 
that this method, used here, with the second order perturbation in the  
wave function could match the exact results to a desirable accuracy 
except in a narrow range of $g_+$ from 1 to 1.3. For this narrow range 
of $g_+$ reasonable results are obtained but one should include higher order 
corrections to obtain results  to match with the exact one. It should be  
mentioned that as one decreases the value of $t$ and move towards the 
antiadiabatic limit the convergence becomes better and the region of 
$g_+$, where the perturbation corrections are appreciable, becomes 
narrower. In that case the perturbation method up to the second order 
correction in the wave function would describe the system for a 
wider range of $g_+$.

\vskip 1.0cm
\begin{center}
{\bf 5. Conclusion}
\end{center}
\vskip 0.3cm

In the present work we develop an analytical perturbation method 
within the modified LF approach to deal with an electron-phonon system 
for the whole range of e-ph coupling strength. This method is 
applicable from the antiadiabatic limit ($t< \omega_0$) to the 
intermediate region of hopping ($t\sim \omega_0$). Considering a two-site 
one electron system we have calculated the single electron energies, 
the ground state oscillator wave functions and the correlation functions 
(up to the second order perturbation correction to the wave function)
and find that the results are in the good agreement with the exact 
results. The perturbation series converges quite rapidly for all values 
of $g_{+}$. 

\newpage
Table I. Variational parameter ($\lambda$), unperturbed single 
electron energy (measured with respect to the bare site energy 
$\epsilon$), second 
order correction and third order correction to the energy for 
different values of the coupling strength ($g_{+}$) for the ground 
state of two-site one polaron problem. 
\vspace{1cm}

\begin{center}
\begin{tabular}{|l|c|c|c|r|}
\hline
{$g_{+}$} & {$\lambda$} & {$E_0^{(0)}-\epsilon$}& {$E_0^{(2)}$}& 
{$E_0^{(3)}$}\\ \hline
{0.2} & {.0628} & {-1.1125} & {-.00007} & {.00000}\\ \hline
{0.5} & {.1619} & {-1.1795} & {-.00275} & {.00001}\\ \hline
{0.8} & {.2771} & {-1.3099} & {-.01767} & {.00062}\\ \hline
{1.0} & {.3763} & {-1.4397} & {-.04564} & {.00522}\\ \hline
{1.1} & {.4421} & {-1.5212} & {-.07266} & {.01408}\\ \hline
{1.2} & {.5363} & {-1.6183} & {-.12168} & {.03805}\\ \hline
{1.3} & {1.0202} & {-1.7489} & {-.28799} & {.09973}\\ \hline
{1.4} & {1.3047} & {-1.9874} & {-.20805} & {.02331}\\ \hline
{1.7} & {1.6875} & {-2.8935} & {-.11787} & {.00094}\\ \hline
{1.9} & {1.8969} & {-3.6108} & {-.09082} & {.00014}\\ \hline
{2.2} & {2.1997} & {-4.8401} & {-.06240} & {.00002}\\ \hline 
\end{tabular}  
\end {center}
\newpage
%\begin{references}

\newpage

Figure captions :

\noindent

FIG. 1. Single electron energies (in units of $\omega_0=1$) as 
a function of the coupling strength ($g_{+}$). Dashed curve: ground 
state, solid curve: first excited state.

\vskip 0.5 cm
\noindent

FIG. 2. Ground state oscillator wave function $G(x)$ as a function 
of x for different values of the coupling strength when the electron 
is located on site 1. Solid: 
$g_{+}=0.1$, short-dashed: $g_{+}=0.7$, long-dashed: $g_{+}=1.3$ and 
dot-dashed: $g_{+}=2.0$.

\vskip 0.5 cm     
\noindent

FIG. 3. Plot of the correlation functions
(a): $\langle n_1u_1 \rangle_{0}$
and (b): $\langle n_1u_2 \rangle_{0}$ versus $g_{+}$. To compare the
results of Ref. (4) we use a unit of
$\frac{1}{2} \sqrt{\frac{\hbar}{M\omega}}$ for the
correlation functions.
 

\begin{thebibliography}{999}
%\begin{center}
%{\bf References }
%\end{center}
%\begin{enumerate}

\bibitem{Hol}{T. Holstein, Ann. Phys. (NY) {\bf 8}, 325 (1959).}

\bibitem{Mig}{A. B. Migdal, Sov. Phys. JETP {\bf 7}, 996 (1958).}

\bibitem{LF}{ L. G. Lang and Yu A. Firsov, Sov. Phys. JETP {\bf 16}, 
1301 (1963).}
 
\bibitem{RT}{J. Ranninger and U. Thibblin, Phys. Rev. B {\bf 45},  
7730 (1992).}

\bibitem{Mar}{F. Marsiglio, Phys. Letts. A {\bf 180}, 280 (1993);  
Physica C {\bf 244}, 21 (1995).}

\bibitem{KR}{V. V. Kabanov and D. K. Ray, Phys. Letts. A {\bf 186}, 
438 (1994).}  

\bibitem{Alex}{A. S. Alexandrov, V. V. Kabanov and D. K. Ray, 
Phys. Rev. B {\bf 49}, 9915 (1994).}

\bibitem{MR}{E. V. L. de Mello and J. Ranninger, Phys. Rev. B {\bf 55}, 
14872 (1997); Phys. Rev. B {\bf 59}, 12135 (1999).}

\bibitem{DC}{A. N. Das and P. Choudhury, Phys. Rev. B {\bf 49}, 
13219 (1994).}

\bibitem{CD}{P. Choudhury and A. N. Das , Phys. Rev. B {\bf 53}, 
3203 (1996).}

\bibitem{DS}{A. N. Das and S. Sil, Phyica C {\bf 207}, 51 (1993); 
J. Phys.: Condens. Matter {\bf 5}, 1 (1993).}

\bibitem{LS}{C. F. Lo and R. Sollie, Phys. Rev. B {\bf 45}, 
7102 (1992).}

\bibitem{FK}{Yu. A. Firsov and E. K. Kudinov, Phys. Solid. State (AIP) 
{\bf 39} (12), 1930 (1997).}

\end{thebibliography}
\end{document}